\documentclass[twocolumn,showpacs,amsmath,amssymb,superscriptaddress]{revtex4}
\usepackage[dvips]{graphicx}
\usepackage{graphics}
\usepackage{dcolumn}
\usepackage{bm}

\begin{document}
\title{Non-Drude  Optical   Conductivity  of  (III,Mn)V  Ferromagnetic
Semiconductors}  
\author{S.-R. Eric  Yang}  
\affiliation{University of
Texas  at  Austin, Physics  Department,  1  University Station  C1600,
Austin  TX  78712-0264}  
\affiliation{  Department of  Physics,  Korea
University,    Seoul    136-701,    Korea}    
\author{Jairo    Sinova}
\affiliation{University  of  Texas at  Austin,  Physics Department,  1
University Station C1600,  Austin TX 78712-0264} 
\author{T. Jungwirth}
\affiliation{University  of  Texas at  Austin,  Physics Department,  1
University   Station  C1600,   Austin  TX   78712-0264}  
\affiliation{ Institute of Physics  ASCR, Cukrovarnick\'a 10, 162 53  Praha 6, Czech
Republic  }  
\author{Y.P. Shim}  
\affiliation{University  of Texas  at
Austin,  Physics Department,  1  University Station  C1600, Austin  TX
78712-0264} \author{A.H. MacDonald} \affiliation{University of Texas at
Austin,  Physics Department,  1  University Station  C1600, Austin  TX
78712-0264} \date{\today}

\begin{abstract}

We present  a numerical model  study of the  zero-temperature infrared
optical properties of  (III,Mn)V diluted magnetic semiconductors.  Our
calculations  demonstrate  the  importance  of treating  disorder  and
interaction effects  simultaneously in modelling  these materials.  We
find that  the conductivity  has no  clear Drude peak,  that it  has a
broadened inter-band peak near 220 meV,  and that oscillator weight is shifted to
higher frequencies  by stronger disorder.   These results are  in good
qualitative agreement  with recent thin  film absorption measurements.
We  use our  numerical  findings to  discuss  the use  of f-sum  rules
evaluated  by   integrating  optical  absorption   data  for  accurate
carrier-density estimates.
\end{abstract}


\maketitle 

\section{ Introduction}

Recent  studies of (III,Mn)V  diluted magnetic  semiconductors (DMSs),
motivated   by  the   discovery  of   carrier-mediated  ferromagnetism
\cite{ohnoprl92}  in  these  materials,  have uncovered  a  wealth  of
interesting  phenomenology involving  an interplay  between collective
electronic  effects, broken  symmetries,  interactions, and  disorder.
The richness of the transport and optical properties of (III,Mn)V DMSs
\cite{ohnoprl92,ohnoapl96}  arises   from  their  strong  valence-band
spin-orbit coupling  and from the sensitivity of  their magnetic state
to  growth  and  annealing  conditions, doping,  and  external  fields
\cite{gallagher,pennstate}.   {\em  In  situ} tunability  of  material
properties   by   applying   gate  voltages   \cite{ohno:nature},   by
illumination\cite{munekata:prl},  or  by  applying  external  magnetic
fields,  suggests  many opportunities  for  exploring  new physics  or
building  new devices  using these  materials.  There  is considerable
evidence that the properties of  these systems can be understood using
a  simplified  effective model  \cite{bookchapter}  in  which the  low
energy  degrees of freedom  are $S=5/2$  Mn$^{2+}$ local  moments, and
holes in the host-semiconductor valence band.  The Hamiltonian of this
effective model should in  general include terms that represent scalar
and exchange  interactions between  the Mn ions  and the  valence band
holes, and  Coulomb interactions among valence band  holes and between
the holes and the Mn ions.  The simplest theoretical approach to these
models,  one that appears  to be  relatively successful  in explaining
many  bulk properties  of strongly  metallic DMS  ferromagnets, either
ignores disorder or treats it perturbatively and approximates exchange
interactions         using         a         mean-field         theory
\cite{dietlsci00,dietlprb01,abolfathprb01,bookchapter}
\footnote{  A large  database with  theory predictions  based  on this
approximation for different host  materials, and different carrier and
Mn densities  can be accessed at  \texttt{ http://unix12.fzu/ms}.}.  We
have recently presented a theory of the intraband optical conductivity
of (III,Mn)V  ferromagnets that  uses such an  approach \cite{sinova}.
In the  present paper we report  a finite-size numerical  study of the
optical  conductivity  of  (Ga,Mn)As  in  which  disorder  is  treated
exactly, enabling us to  address weakly metallic or insulating systems
and to  understand some  of the qualitative  strong-scattering effects
that are omitted in the simpler theory.

Our interest  in this property of  DMS ferromagnets is  motivated by a
series      of     recent      optical      absorption     experiments
\cite{nagai,Katsumoto,HirakawaGaAs,SanDiego}.      The    experiments,
performed  in thin  film geometries  in both  metallic  and insulating
samples (as  judged by the  temperature dependence of  the resistivity
below $T_c$), exhibit several  common features: (a) non-Drude behavior
in which  the conductivity increases with increasing  frequency in the
interval between  $0$ meV and $220$  meV, (b) a  broad absorption peak
near $220-260$  meV that becomes  stronger as the samples  are cooled,
and (c) a broad featureless absorption between the peak energy and the
effective  band gap  energy  which  tends to  increase  at the  higher
frequencies.    The   $220$   meV    peak   can   be   attributed   to
inter-valence-band  transitions \cite{sinova}, to  transitions between
the semiconductor valence band states  and a Mn induced impurity band,
or   to    a   combination   of    these   contributions.    Dynamical
mean-field-theory studies  \cite{Hwang}, for a  single-band model that
neglect the  spin-orbit coupling and  the heavy-light degeneracy  of a
III-V   semiconductor  valence   band,   demonstrate  that   non-Drude
impurity-band  related peaks  in the  frequency-dependent conductivity
occur  in  DMS  ferromagnet  models  when  the  strength  of  exchange
interaction  coupling is comparable  to the  valence band  width.  The
conductivities predicted by this model are, however, inconsistent with
experiment  in temperature  trend and  appear not  to be  in  the same
regime as  experimental samples.  On  the other hand  we \cite{sinova}
demonstrated in earlier  work that peaks in the  conductivity can also
occur  in  the  weak-coupling  regime,  provided  that  the  realistic
multi-band character  of the valence  bands is acknowledged.   In this
approach  the  $220$ meV  peak  is  due  to heavy-hole  to  light-hole
inter-valence-band transitions and  the gradual increase in absorption
observed  in  the experiments  \cite{Katsumoto}  at frequencies  above
$500$ meV is ascribed to  transitions to the bands split-off by strong
spin-orbit interactions.  Although this  theory could account for many
overall  features of  the  measured optical  conductivity in  metallic
samples,  one prominent feature  was at  odds with  experimental data,
namely the relative magnitudes of  the $\omega \to 0$ conductivity and
the $220$ meV conductivity  peak.  The numerical calculations reported
on  in   this  paper   suggest  that  this   discrepancy  is   due  to
multiple-scattering   effects    that   suppress   the   low-frequency
conductivity.

We  restrict our attention  in this  paper to  the $T=0$  limit, which
allows  us to neglect  scattering of  valence band  quasiparticles off
thermal fluctuations  in the  Mn ion spin  orientations.  We  assume a
simple  ground  state  in  which  the  Mn  spins  have  parallel  spin
orientations
\footnote{The combined effect  of disorder and spin-orbit interactions
implies  that the classical  ground state  of a  (III,Mn)V ferromagnet
cannot  have perfectly  parallel Mn  spin orientations.   Disorder can
also lead to frustration in the carrier-mediated Mn spin interactions,
and to  complex ground  states in which  spins are not  aligned.  Both
effects  become less  important for  more strongly  metallic (III,Mn)V
ferromagnets.   See Ref.  \onlinecite{janko,schliemann}.}.   Sources of
disorder  that  are  known   to  have  some  importance  in  (III,Mn)V
ferromagnets include randomness in the  placement of Mn ions on cation
sites of the host lattice, random placement of Mn ions at interstitial
sites, and randomly located antisite defects in which group V elements
are  placed at  group III  element lattice  sites.  In  the low-energy
effective  model, Mn  ions at  cation sites  have charge  $Q=-e$ while
interstitial Mn and antisite defects have $Q=2e$.
Coulomb  and exchange interactions  with the  Mn ions  have comparable
importance \cite{sinova} for the  scattering of valence band holes and
must  both  be  included in  the  theory.   In  the approach  of  Ref.
\onlinecite{sinova} these sources  of disorder scattering were treated
perturbatively using a Born approximation, evaluating the valence band
quasiparticle  lifetimes using Fermi's  Golden rule  \cite{tomas}.  In
this  paper we  find the  $T=0$ quasiparticle  states  for finite-size
systems  by solving  the  Schr\"{o}dinger equation.   Unlike the  Born
approximation theory, this approach  does not assume metallic behavior
and  remains  valid  at  low  Mn concentrations  when  \cite{eric}  Mn
acceptor impurity bands emerge.  We  find that corrections to the Born
approximation theory  are important at  a quantitative level  even for
strongly metallic high Mn concentration samples, and that they explain
the  property  that  the   low-frequency  and  dc  conductivities  are
suppressed compared to their values in the Born approximation theory.
We  also  find  that  the  f-sum rule,  obtained  by  integrating  the
conductivity over frequency, differs from its Born approximation value
by less than 10\% for typical metallic carrier densities.

This  paper is  organized as  follows.  Section  II defines  the model
Hamiltonian we use to describe  the valence band system in a (III,Mn)V
ferromagnet which  adds exchange and  Coulomb interactions to  the six
band  ${\bf  k\cdot p}$  envelope  function  Hamiltonian  of the  host
(III,V)  semiconductor.    Section  III  describes   technical  issues
associated with the evaluation of Kubo linear-response-theory formula
for  the  finite-size system  optical  conductivity.   In  Sec. IV  we
discuss the  numerical results we  have obtained for different  Mn and
carrier  density  regimes, which  are  in  qualitative agreement  with
experiment  and discuss  the degree  to which  aspects of  the extreme
impurity-band \cite{Hwang}  and Golden-rule \cite{sinova,tomas} limits
are reflected in the exact results.  We also emphasize the feasibility
of using the f-sum rule to extract reasonably accurate measurements of
the  carrier concentration  \cite{sinova}, circumventing  the inherent
uncertainties  in Hall  measurements in  systems with  large anomalous
Hall effect coefficients.

\section{Model Hamiltonian} 

We  approximate the host  semiconductor valence  band by  its six-band
${\bf k}\cdot{\bf p}$ Kohn-Luttinger model.  This simplification takes
advantage of the small number of holes per atom in these ferromagnets.
The band electrons interact via Coulomb and exchange interactions with
randomly distributed  Mn spins and with  randomly distributed antisite
defects via  Coulomb interactions alone; we do  not explicitly account
for the possible role  of Mn interstitials \cite{Yu}.  The requirement
of overall charge neutrality implies  that the density of valence band
holes ($p$) is related to the density of Mn ($n_{Mn}$) and the density
of  antisites ($n_{AS}$)  by $p=  n_{Mn} -  2 n_{AS}$.   In  MBE grown
(III,Mn)V  ferromagnets,  it  is   often  a  challenge  to  accurately
determine the density  of holes, and even the density  of Mn ions that
have been incorporated into the ferromagnetic state, although progress
is being  made.  It  appears at present  that any  annealing procedure
\cite{pennstate,schiffer}  that  leads  to higher  carrier  densities,
inevitably leads to  higher ferromagnetic transition temperatures, and
to     much    higher     conductivities     \cite{gallagher}.     For
Ga$_{0.95}$Mn$_{0.05}$As for  example, it  is now possible  to prepare
samples with  $T=0$ dc  conductivities that are  close to an  order of
magnitude  larger  than  in  the  samples for  which  ac  conductivity
measurements have so  far been carried out.  Although  we believe that
much of the variabilty in  physical properties can be accounted for in
terms of variations  in the fraction of Mn  ions that are incorporated
and, more  importantly, variations in the carrier  density, it appears
likely that a part of  this variation is still associated with sources
of unintended disorder  that are not present in  our model.  We expect
that  as  unintended disorder  is  reduced,  the  properties of  these
ferromagnets will be more  completely determined by the two parameters
of our model calculations, $p$ and $n_{Mn}$
\footnote{There is a broad consensus  that variation in the density of
Mn interstitials is  responsible for most of the  variation of carrier
density  and magnetically  active Mn  density that  occurs  when these
materials are annealed.  See for example Ref. \onlinecite{Yu}.}.

Below we  will evaluate the frequency-dependent  conductivity by using
the Kubo formula description for the linear response of quasiparticles
to an  electromagnetic field.  Coulomb interactions  play an essential
role in  determining the  quasiparticle states and  interactions among
the  holes  cannot  be  ignored.   In  this paper  we  use  a  Hartree
mean-field approximation  for the quasiparticles,  which is consistent
with  the  standard  bubble  diagram  approximation  we  use  for  the
conductivity.

The  single particle  part  of the  quasiparticle  Hamiltonian may  be
written  as  the  sum  of   host  crystal  band  and  disorder  pieces
$\hat{H}^0=\hat{H}^h+\hat{H}^D$:  The disorder  Hamiltonian  has three
contributions
$\hat{H}^D=\hat{H}^{K.ex.}+\hat{H}^{Mn-h}+\hat{H}^{As-h}$.
$\hat{H}^{Mn-h}$ describes the  attractive Coulomb interaction between
the   ionized   $Mn^{2+}$   acceptor   and  a   valence   band   hole,
$\hat{H}^{K.ex.}$  describes the  local  kinetic-exchange interactions
between   valence   band   carriers   and  the   Mn   moments,   while
$\hat{H}^{As-h}$ describes the repulsive Coulomb interaction between a
hole  carrier  and  ionized  anti-site defects.   The  single-particle
envelope function model Hamiltonian then reads:
\begin{widetext}
\begin{equation}
\hat{H}^0=\hat{H}^h+\sum_{I=1}^{N_{Mn}}
\vec{S}_I\cdot\hat{\vec{s}}J(\vec{r}-\vec{R_I})+
\sum_{I=1}^{N_{Mn}}(-\frac{e^2}{\epsilon|\vec{r}-\vec{R_I}|}
-V_0e^{-|\vec{r}-\vec{R_I}|^2/r_0^2})\hat{I}
+\sum_{K=1}^{N_{As}}\frac{2e^2}{\epsilon|\vec{r}-\vec{R_K}|}\hat{I}
\label{charge}
\end{equation}
\end{widetext}
where $J(\vec{r})=(J_{pd})/((2\pi  a_0^2)^{3/2}) e^{-r^2/2a_0^2}$, and
$\hat{\vec{s}}=(\hat{s_x},\hat{s_y},\hat{s_z})$                   where
$\hat{s}_{x,y,z}$  are the $6\times6$  matrices which  describe hole
spins  in  the  representation   of  the  six  band  envelope-function
Hamiltonian,  and $\hat  I$  is  a $6\times6$  unit  matrix.  In  this
Hamiltonian, $I$ labels Mn sites, $K$ labels As sites, and $\vec{S}_I$
stands  for a  Mn spin  with quantum  number $S=5/2$.   The Mn  and As
positions  are  denoted  by  $\vec{R_I}$ and  $\vec{R_K}$.   The  term
proportional to  $V_0$ is a  central cell correction  \cite{bhatt} to
the Coulomb attraction between holes and Mn acceptors that is known to
be required  to recover experimental values for  the isolated acceptor
energy.
The quantities $N_{Mn}$ and $N_{As}$ are the number of Mn and As atoms, which 
we distributed randomly inside cubic simulation cells of side $L$.
Charge neutrality in our simulation cell
means that the number of quasiparticle levels occupied by holes is 
$N_h = N_{Mn} - 2 N_{As}$. 
The values chosen for the phenomenological parameters that appear in this equation
are $J_{pd}=0.05$eV nm$^3$, $a_0=0.3$ nm,
$\epsilon=10.9$, $V_0=2.5eV$ and $r_0=0.259$ nm.

The host  band part of the  Hamiltonian is described via  the six band
Kohn-Luttinger  model,  with  wavevectors  $\vec{k}$ in  the  envelope
function   replaced   by   $k_a=\frac{1}{i}\nabla_a$   for
$a=x,y,z$  to  account  for  the inhomogeneity  induced  by  disorder.
Choosing the  angular momentum quantization direction to  be along $
z$-axis, and ordering  the $j=3/2$  and $j=1/2$  basis functions
according to the  list ($-3/2,1/2,-1/2,-3/2;1/2,-1/2$), the Luttinger
Hamiltonian $H^L$ has the form \cite{abolfathprb01}:
\begin{equation}
H^L = \left(\begin{array}{cccccc} {\cal H}_{hh} & -c & -b &
\multicolumn{1}{c|}{0} & \frac{b}{\sqrt{2}} & c\sqrt{2}\\ -c^* & {\cal H}
_{lh}
& 0 & \multicolumn{1}{c|}{b} & -\frac{b^*\sqrt{3}}{\sqrt{2}} & -d\\ -b^*
& 0 &
{\cal H}_{lh} & \multicolumn{1}{c|}{-c} &   d & -\frac{b\sqrt{3}}{\sqrt{2
}} \\
0 & b^* & -c^* & \multicolumn{1}{c|}{{\cal H}_{hh}} &  -c^*\sqrt{2} &
\frac{b^*}{\sqrt{2}}\\ \cline{1-4} \frac{b^*}{\sqrt{2}} &
-\frac{b\sqrt{3}}{\sqrt{2}} & d^* & -c\sqrt{2} & {\cal H}_{so} & 0\\
c^*\sqrt{2} & -d^* & -\frac{b^*\sqrt{3}}{\sqrt{2}} & \frac{b}{\sqrt{2}} &
 0 &
{\cal H}_{so}\\
\end{array}\right)
\label{hl}
\end{equation}
In the matrix (\ref{hl}) we have highlighted the $j=3/2$ sector.
The Kohn-Luttinger eigenenergies are measured down from the top of the valence
band, i.e. they are hole energies.  For completeness we list the expressions which define
the quantities that appear in $H^L$:
\begin{eqnarray}
{\cal H}_{hh} &=& \frac{\hbar^2}{2m}\big[(\gamma_1 + \gamma_2)(k_x^2+k_y^
2) +
(\gamma_1 - 2\gamma_2)k_z^2 \nonumber 
\end{eqnarray}
\begin{eqnarray}
{\cal H}_{lh} &=&
\frac{\hbar^2}{2m}\big[(\gamma_1 - \gamma_2)(k_x^2+k_y^2) + (\gamma_1 +
2\gamma_2)k_z^2 \nonumber \\ {\cal H}_{so} &=&
\frac{\hbar^2}{2m}\gamma_1(k_x^2+k_y^2+k_z^2) + \Delta_{so} \nonumber
\end{eqnarray}
\begin{eqnarray}
b &=&
\frac{\sqrt{3}\hbar^2}{m} \gamma_3 k_z (k_x - i k_y) \nonumber \\ c &=&
\frac{\sqrt{3}\hbar^2}{2m}\big[\gamma_2(k_x^2 - k_y^2) - 2i\gamma_3 k_x
k_y\big] \nonumber 
\end{eqnarray}
\begin{eqnarray}
d &=&
-\frac{\sqrt{2}\hbar^2}{2m}\gamma_2\big[2k_z^2-(k_x^2 + k_y^2 )\big]\; .
\label{lutpar}
\end{eqnarray}
with $\gamma_1=6.98$, $\gamma_2=2.06$ and $\gamma_3=2.93$.

The Hartree potential due to interactions among holes must also be included in the 
single-particle Hamiltonian for quasiparticle states
since it captures the screening of Coulomb interactions with impurities
due to the build up or depletion of charge.  
The total one-particle Hamiltonian is $ \hat{H}=\hat{H}^0+\hat{V}^H$, where
\begin{eqnarray}
V^H_{\vec{k}j,\vec{k'}j'}=\delta_{j,j'}\sum_{j"}
\sum_{\vec{p}}\frac{4\pi e^2}{\epsilon|\vec{k}-\vec{k'}|^2}\rho_{\vec{p}j",
\vec{k'}+\vec{p}+\vec{k}j"},
\end{eqnarray}
and the density matrix
\begin{eqnarray}
\rho_{\vec{k}j,\vec{k'}j'}
&=&\sum_{\alpha}f_{\alpha}<\Psi_{\vec{k}j}|\alpha><\alpha|\Psi_{\vec{k'}j'}>\nonumber\\
&=&
\sum_{\alpha}f_{\alpha}
c_{\vec{k}j}^{(\alpha)}c_{\vec{k'}j'}^{(\alpha)*},
\end{eqnarray}
with $f_{\alpha}=1 (0)$ for an occupied (unoccupied) state.
Here $\Psi_{\vec{k}j}(\hat{r})=e^{i\hat{k}\cdot\hat{r}}u_{\vec{k}j}(\vec{r})$, where 
$u_{\vec{k}j}(\vec{r})$ are the Bloch functions.
The Hartree potential must be determined by solving self-consistently for the density matrix.
We have diagonalized this Hamiltonian using a plane-wave representation for the 
envelope functions of $\hat{H}^0$ and applying periodic boundary conditions in a cube of size $L$,
which limits wavevectors to the discrete set $\vec{k}=(n_x, n_y, n_z)(2\pi/L)$,
where $n_x,n_y,n_z$ are integers.  
For different system sizes we choose different
maximum values of $n_{x,y,z}$, denoted by $n_m$, so that the maximum wavevector $k_{max}$ is
held fixed.  For the calculations presented here $k_{max}=\pi$ nm$^{-1}$.

\section {Kubo formula for the Optical conductivity}
We evaluate the real part of the frequency dependent conductivity from the Kubo formula 
expression which relates it to the quasiparticle eigenvectors
$|\alpha> $ and eigenvalues $E_{\alpha} $.
\begin{eqnarray}
\sigma_{ab}(\omega)&=&\frac{\pi e^2}{m^2 L^3 \omega}\sum_{\alpha,\beta}(f_{\alpha}-f_{\beta})
\langle\alpha|\hat{p}_a|\beta\rangle\langle\beta|\hat{p}_b|\alpha\rangle \\&&\times
\delta(\hbar \omega-E_{\beta}+E_{\alpha}).\nonumber
\end{eqnarray}
Matrix elements of the momentum operator, 
$\hat{p}_a=\frac{\hbar}{i}\nabla_a$ 
with $a=x,y,z$, are given by
\begin{eqnarray}
<\alpha|\hat{p}_a|\beta>&=&\sum_{j,j'}\sum_{\vec{k},\vec{k'}}
c_{\vec{k}j}^{(\alpha)*}c_{\vec{k'}j'}^{(\beta)}
<\Psi_{\vec{k}j}|\hat{p}_a|\Psi_{\vec{k'}j'}>\nonumber\\
&=&\sum_{\vec{k},j}c_{\vec{k}j}^{(\alpha)*}c_{\vec{k}j'}^{(\beta)}
\frac{m}{\hbar}\frac{\partial H^L_{j,j'}}{\partial k_a}.
\end{eqnarray}
At zero temperature and positive $\omega$ the difference  $(f_{\alpha}-f_{\beta})$ restricts
$\alpha$ to occupied quasiparticle states and $\beta$ to empty quasiparticle states.
For finite-size systems the real part of the conductivity 
is more reliably evaluated by broadening the $\delta$ function \cite{Imrybook}
\begin{eqnarray}
C^{ab}_R(\omega)&=& \mathrm{Re}\Biggl[\frac{\hbar e^2}{m^2L^3 }\sum_{\alpha,\beta}
\frac{f_{\alpha}-f_{\beta}}{E_{\beta}-E_{\alpha}}
<\alpha|\hat{p}_a|\beta>\nonumber\\&&\times
<\beta|\hat{p}_b|\alpha> 
\frac{ \gamma}{ (\hbar\omega-E_{\beta}+E_{\alpha})^2 + \gamma^2}\Biggr],
\end{eqnarray}
with the level  broadening, $\gamma$ , being of the  order of the mean
level spacing  $\sim E_F/N_h\sim 1/L^3$.   In the thin film  limit the
real  part  of the  conductivity  is  proportional  to the  absorption
coefficient  in  transmission  experiments,  $\tilde{\alpha}(\omega)=\frac{
8\pi  Re[\sigma(\omega)] }{c(1+n)}$  with  $n=\sqrt{\epsilon_0}$ being  the
zero frequency limit index of refraction of the thin film material and
$\epsilon_0$ the dimensionless  dielectric constant.  This formula for
$\tilde{\alpha}(\omega)$ is  most accurate in the  far infrared regime
or  in  the thin  film  limit  since  at higher  frequencies  multiple
scattering from the  interfaces must be taken into  account.  This can
be easily done using the standard electromagnetic description of light
waves propagating through a  medium, but the simple proportionality is
lost.

\section{Results}

\subsection{Optical conductivity of parabolic band with disorder}

It is useful for pedagogical purposes to first present frequency-dependent conductivities 
calculated with this approach for the case where the 
6-band Luttinger model is replaced by a parabolic band model with 
a mass equal to the heavy hole mass of GaAs.
In this way we can partially separate the effects of strong disorder 
on the optical conductivity from the complications associated with the 
heavy, light, and split-off valence bands.
In the calculation all Mn spins point along the z-axis so that 
spin up and down single-particle Hamiltonians decouple.
For each spin the Hamiltonian is identical to that of  Eq.~1 
except that $\hat{H}^L$ is replaced by the usual kinetic 
term $-\hbar^2\nabla^2/2m_{hh}$ and the $6\times 6 $ identity matrix is replaced by unity.
The following parameters were used in the calculation of the conductivity: 
p=0.33~nm$^{-3}$, $n_{Mn}=1$~nm$^{-3}$ ($x\approx 4.5$\%), $L=8$~nm,  
and $n_m=4$.  
The results presented below were obtained by averaging over $N_D=10$ realizations 
of the disorder potential. 
\begin{figure}
\includegraphics[width=3.4in]{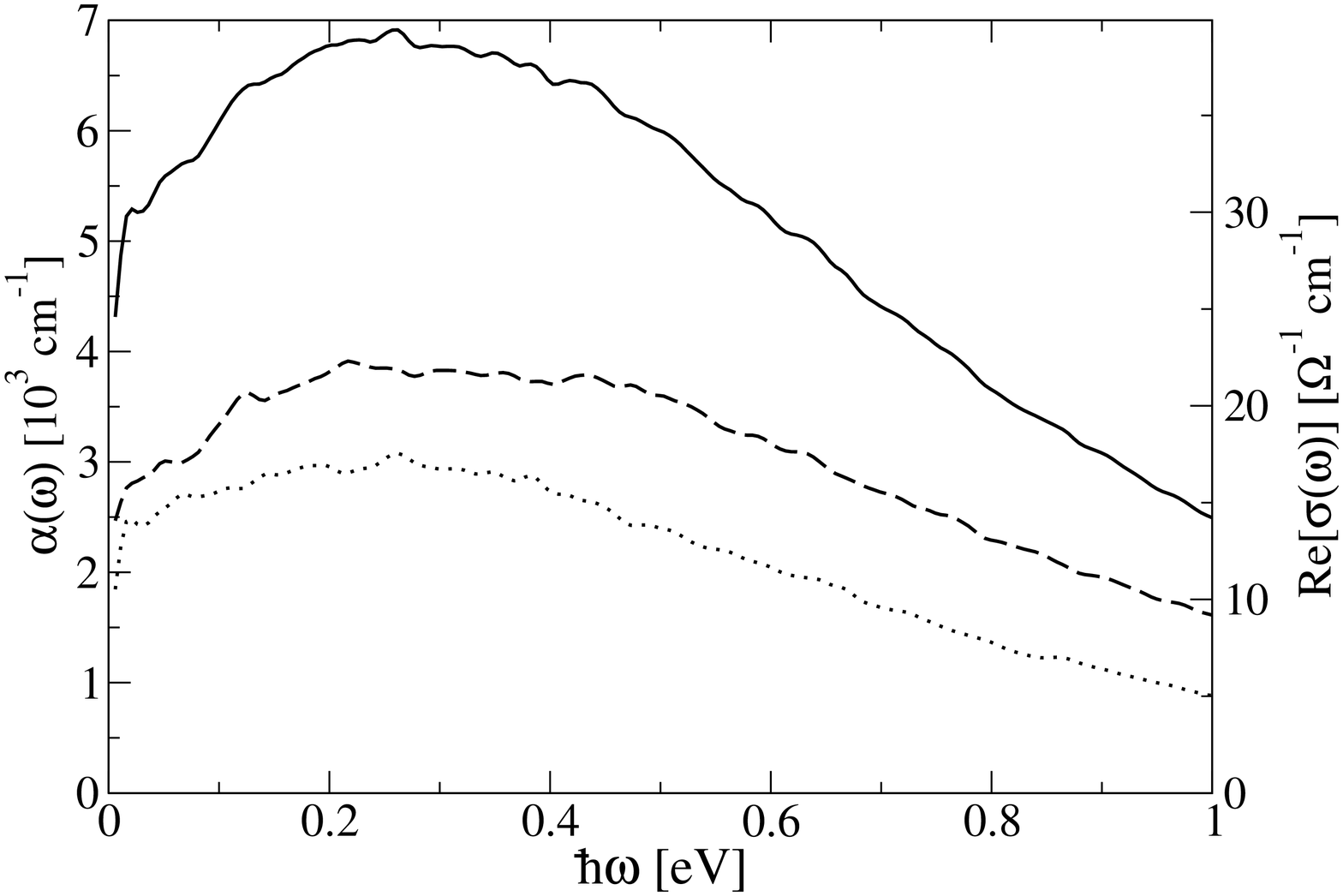}
\caption{\label{figure1}  Absorption   and  conductivity  of  minority
(dotted line) and majority (dashed  line) carriers in a parabolic band
approximation.   Solid line  is  the total  absorption.  Here  
$p=0.33$~nm$^{-3}$, $n_{Mn}=1$~nm$^{-3}$ ($x\approx 4.5$\%).}
\end{figure}

Fig.~\ref{figure1}   showss     conductivities  and   absorption
coefficients  we have  evaluated for  minority and  majority carriers.
Even  for  this simple  band  model  the  conductivity displays  clear
non-Drude  behavior, increasing  as  a function  of  frequency at  low
frequencies.   We believe that  the non-Drude  behavior below  200 meV
reflects   multiple  scattering   effects   that  enhance   high-angle
scattering  and  would  lead  to  localization if  the  disorder  were
stronger.   These effects  are explicitly  neglected in  the Boltzmann
transport theory which leads to the Drude formula.  The f-sum rule for
the  conductivity calculated  by integrating  these  numerical results
over frequency  deviates from the  parabolic band model value  by less
than $10 \%$, demonstrating that  the errors induced by constructing a
smooth  conductivity  curve using  broadened  $\delta$- functions  are
under reasonable control.  The f-sum rule is more accurately satisfied
for  larger system  sizes, for  which we  are able  to  choose smaller
values of  the broadening parameter $\gamma$.   The non-Drude behavior
we  find here  does capture  one aspect  of the  experimental results.
However  the overall  conductivity  magnitude of  this  model is  much
smaller than in  experiment.  Evidently the mixing of  heavy and light
hole bands leads to quasiparticles with larger velocities.

\subsection{Optical conductivity of Luttinger-Kohn Hamiltonian with disorder}
Mixing of heavy- and light-hole  bands and its interplay with magnetic
order and scattering disorder can  be investigated by solving the full
Hamiltonian  $H=\hat{H}^0+\hat{V}^H$.  This  approach treats  the band
structure  realistically and the  randomly located  charged impurities
without approximation.
\begin{figure}
\center \includegraphics[width=3.3in]{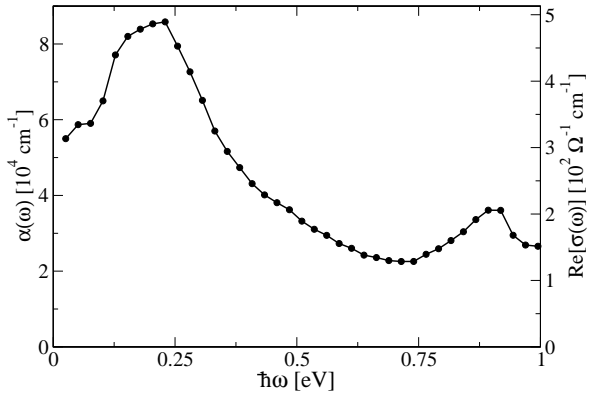}
\caption{Absorption  and conductivity  of a  metallic  sample computed
using a Luttinger Kohn  Hamiltonian with disorder.  Nominal parameters
same as in Fig. \ref{figure1}.}
\label{figure2}
\end{figure}
Fig.  \ref{figure2} shows the frequency-dependent conductivity we have
evaluated  for   $p=0.33$~nm$^{-3}$,  $n_{Mn}=1$~nm$^{-3}$,  $L=6$~nm,
$N_D=4$ and  $n_m=3$.  For  $\hbar\omega<200$~meV we observe  that the
conductivity increases with frequency,  unlike the Drude formula case.
The increase is  actually sharper than in the  parabolic band model, a
property  that  we  believe  reflects the  additive  contributions  of
non-Drude  intra-band   absorption  and  broadened  inter-valence-band
absorption.  There  are significant differences  between these results
and those of  the Born approximation (see Fig.1  of Sinova \textit{ et
al.}  \cite{sinova}) calculated for the same model parameters.
In  our calculation  the  maximum  value of  the  absorption is  about
90000~cm$^{-1}$ while in the calculation of Sinova \textit{ et al.} it
is about 110000~cm$^{-1}$.   We believe that this reduction  is due to
the enhanced backscattering effects  captured by the exact calculation
and also  due to a transfer  of spectral weight out  of the intra-band
contribution.   The  peak   near  $\hbar\omega\sim  0.9$~eV  in  these
calculations is due to heavy-hole to split-off band transitions and is
more pronounced here than in the Born approximation calculations.  The
presence  of this  peak  at rather  large  frequencies emphasizes  one
weakness of  the present calculation, namely that  it neglects valence
to conduction band transitions.
It  will be  interesting  to  see whether  or  not future  experiments
confirm  our prediction  of  a peak  in  the absorption  due to  these
transitions that lies at high frequencies, but still clearly below the
onset of valence-to-conduction band transitions.
\begin{figure}
\center \includegraphics[width=3.3in]{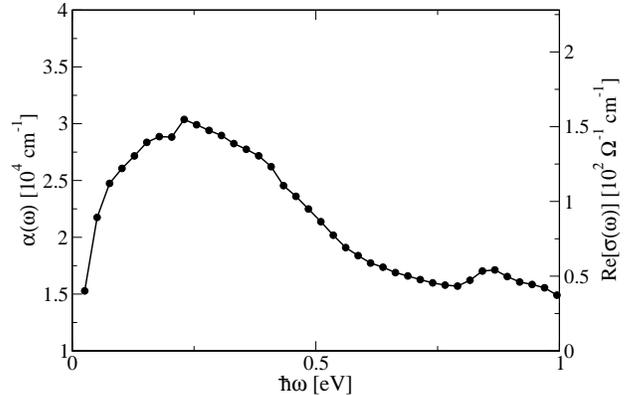}
\caption{Absorption and  conductivity of a  disordered sample computed
using   a   Luttinger    Kohn   Hamiltonian.    The   parameters   are
$p=0.2$~nm$^{-3}$,  $n_{Mn}=0.88$~nm$^{-3}$,   $L=6$~nm,  $N_D=4$  and
$n_m=3$.}
\label{figure3}
\end{figure}

The results  shown in Fig.~\ref{figure2}  are for parameters  that are
typical for high  $T_c$ (Ga,Mn)As ferromagnets.  The dc  limits of the
conductivities that  we calculate are substantially  higher than those
of the  samples for which $\sigma(\omega)$ values  have been published
at present, but  comparable to or smaller than  those measured for the
highest  conductivity   (Ga,Mn)As  ferromagnets  currently  available.
Fig.~\ref{figure3}  displays the  absorption curves  calculated  for a
more disordered system
\footnote{ For  carrier density drops below $\sim  0.2 {\rm nm}^{-3}$,
disorder  localizes  the  valence  band electrons  \cite{eric}.   More
strongly localized valence band  holes have more heavy-hole character,
causing the the optical mass  to increase in comparison to values that
are  appropriate   in  the  metallic  limit.}.    The  parameters  are
$p=0.2$~nm$^{-3}$,  $n_{Mn}=0.88$~nm$^{-3}$,   $L=6$~nm,  $N_D=4$  and
$n_m=3$.   The   conductivity  is  substantially   suppressed  at  all
frequencies, but  most strongly at low  frequencies.  It
again  has  non-Drude  behavior  at low-frequencies,  increasing  with
increasing frequency over the interval between 0~meV and 220~meV.  The
broad heavy-to-light  absorption peak near 220~meV is  still clear but
the  heavy-hole  to  split-off  band  peak is  less  visible,  perhaps
explaining   the  fact   that   it  has   not   yet  been   identified
experimentally.
Hirakawa \textit{et al.} report  that the maximum value for absorption
in  the more disordered  sample they  study is  about 20000~cm$^{-1}$,
which has the same order of  magnitude as the value calculated for the
parameters of Figure \ref{figure3}.

These results  are qualitatively  consistent with overall  features of
the  measured optical  conductivity, including  the  overall magnitude
\cite{nagai,SanDiego,Katsumoto}.    The  accuracy  of   our  numerical
results is  limited somewhat  by the wavevector  cutoffs that  we must
invoke in  order to make the numerical  calculations manageable.  This
is especially  troublesome in estimating the  importance of heavy-hole
to  split-off band  transitions.  The  gradual increase  in absorption
beyond 500~meV  seen in experiment may reflect  these transitions.  We
expect that allowing more  wavevectors in our calculations would shift
oscillator strength toward higher energies and produce a broader peak,
in better agreement with experiments.
(Increasing $n_m$ from  our current value, 3, by  just one changes the
dimension of the Hamiltonian  matrix from 2058 to 4374,  and the resulting
self-consistent computation becomes  quite substantial.)  On the other
hand, our  investigation of  parabolic band case,  where $n_m$  can be
increased to  4, suggests  that system size  dependence of  low energy
non-Drude part of absorption is negligible for $n_m$ bigger than 3.

\section{Conclusions}

We  have presented  a  non-perturbative self-consistent  study of  the
optical properties of diluted (III,Mn)V DMS which treats both disorder
and interactions  on equal footing. Our calculations  for the diagonal
ac-conductivity   are  in   good  qualitative   or  semi-quantitative
agreement with optical absorption experiments in thin film geometries.
Several  of the non-Drude  behaviors observed  in the  experiments are
naturally  accounted for  in this  approach which  is able  to capture
multiple scattering effects that are beyond the scope of perturbative
Born  approximation approaches  \cite{sinova}.  Our  calculations show
that spin-orbit coupling of valence band holes must be included in the
calculation in  order to account  for the qualitative features  of the
experiments.
It is  possible that features associated with  heavy-hole to split-off
band  absorption will  emerge more  clearly near  0.9~eV  when $\sigma
(\omega)$ measurements  are performed  on the most  metallic (Ga,Mn)As
ferromagnets currently available.

We have used these calculations to test the efficacy of the f-sum rule 
we proposed in earlier work for carrier density measurements.  
In applying the f-sum rule to the present system a complication 
arises, compared to microscopic and parabolic band cases, in that 
the integrated conductivity depends on the character of the occupied
quasiparticle states.  In previous work \cite{sinova} we estimated
that for carrier densities in the range that is interesting for 
(Ga,Mn)As ferromagnets, the effective mass that should be used in the 
f-sum rule is $\sim 0.24 {m_e}$, intermediate between heavy and 
light-hole masses.  This estimate was based on calculations of the 
occupied quasiparticle states that 
treated disorder perturbatively in the Born approximation.
We have found by integrating the optical conductivities 
evaluated here over frequency that this optical effective mass estimate 
is reliable at the $\sim 10\%$ level, when disorder is treated exactly.
Uncertainties introduced by our finite momentum cut-off do not allow
us to estimate the value or the variability of this mass more accurately. 
Nevertheless, this level of accuracy should be comparable with 
what is possible from Hall effect measurements which are complicated by 
the strong anomalous Hall effect in these ferromagnets.

\begin{acknowledgments}
This work was supported in part by KOSEF Quantum-functional 
Semiconductor Research Center at Dongkuk University, the Welch Foundation, 
DARPA, the Grant Agency of the Czech Republic
under grant 202/02/0912, and 
the Ministry of Education of the Czech Republic
under grant OC P5.10.
\end{acknowledgments}


\end{document}